\documentclass{article}
\usepackage{arxiv}
\usepackage{graphicx,amsmath,amsfonts,amscd,amssymb,pstricks}
\newcommand{\Year}[1] {(#1)}
\newcommand{\Page}[1] {p.~#1}
\newcommand{\etal} {{\sl et al.}}
\newcommand{\REVIEW}[4] {{\sl #1} {\bf #2} #4 (#3)}
\newcommand{\Publ}[1] {(#1)}
\newcommand{\Book}[1] {{\sl #1}}
\newcommand{\Editor}[1] {ed. by #1}

\title{Numerical simulations of electron-cloud build up in circular accelerators in the presence of multimode-distribution beams}  
\author{X.~Cui\\ 
Institute of High Energy Physics, Beijing, China\\
\And
S.~Gilardoni \\
CERN, CH-1211 Geneva 23, Switzerland \\
\And
M.~Giovannozzi\thanks{Corresponding author: massimo.giovannozzi@cern.ch} \\
CERN, CH-1211 Geneva 23, Switzerland \\
\And 
G.~Iadarola \\
CERN, CH-1211 Geneva 23, Switzerland}
\begin{document}
\maketitle

\begin{abstract}
Electron cloud effects have become one of the main performance limitations for circular particle accelerators operating with positively-charged beams. Among other machines worldwide, the CERN Super Proton Synchrotron (SPS), as well as the Large Hadron Collider (LHC) are affected by these phenomena. Intense efforts have been devoted in recent years to improve the understanding of electron cloud (EC) generation with the aim of finding efficient mitigation measures. In a different domain of accelerator physics, non-linear resonances in the transverse phase space have been proposed as novel means of manipulating charged particle beams. While the original goal was to perform multi-turn extraction from the CERN Proton Synchrotron (PS), several other applications have been proposed. In this paper, the study of EC generation in the presence of charged particle beams with multimode horizontal distribution is presented. Such a peculiar distribution can be generated by different approaches, one of which consists in splitting the initial Gaussian beam distribution by crossing a non-linear resonance. In this paper, the outcome of detailed numerical simulations is presented and discussed.
\end{abstract}
%
%
%
\section{Introduction}
\begin{figure}[htb]
   \centering
   \includegraphics[trim= 5mm 18mm 5mm 22mm,width=\linewidth,clip=]{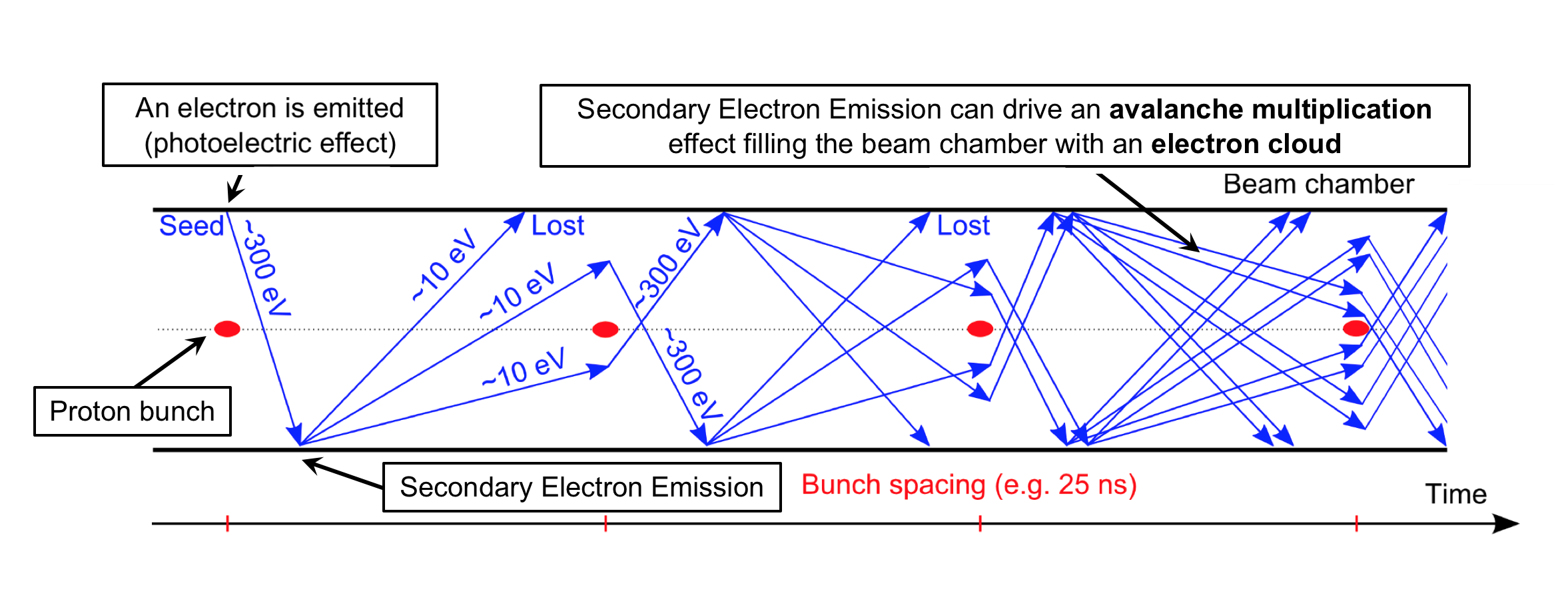}
   \caption{Schematic illustration of the EC buildup process.}
   \label{fig:sketch}
\end{figure}

Over the last five decades, EC effects have been observed in several circular accelerators operating with positively-charged particles~{\cite{izawa,cimino, zimmermann}}. The mechanism leading to the formation of an EC in the beam chamber of a particle accelerator is illustrated schematically in Fig.~\ref{fig:sketch}~{\cite{ohmi,furmanPAC, giannithesis}}. Primary or seed electrons can be generated by a bunch passage due to the ionisation of the residual gas or to photo-emission from the chamber's wall induced by the beam synchrotron radiation.

These electrons can be accelerated by the electric field of the beam, typically to energies in the order of hundreds of electronvolts and when they impact on the walls, secondary electron emission might occur and multiple lower-energy electrons would be emitted. These secondary electrons have even lower kinetic energy ($\sim 10$~eV), hence, in case of impact on the wall, there is a high probability of absorption without generating any further electrons.

Provided that the spacing between subsequent bunches is sufficiently short, the electrons can be accelerated by the following bunch passage before impacting on the wall, which strongly increases the probability of generating more electrons. In the case of a long bunch train, this can lead to an avalanche effect, the so-called multipacting regime, which is responsible for the formation of a dense EC in the chamber. This mechanism increases the electron density for the bunches at the tail of {the} bunch train.

EC can induce unwanted effects on the circulating beam, such as transverse instabilities, transverse emittance blow-up, and particle losses. Moreover, vacuum degradation due to electron-stimulated desorption~\cite{1288839} and power deposition on the chamber's walls can be observed. All these phenomena can lead to severe performance limitations, which explains the intense efforts devoted to the study of EC-related phenomena and possible mitigation measures. The CERN LHC~\cite{LHCDR} and its high-luminosity upgrade~\cite{HL-LHC} are excellent examples of accelerators whose performance can be strongly affected by EC.

Vacuum degradation due to electron-stimulated desorption can pose different problems, i.e increased background in collider experimental regions and risk of breakdown in high-voltage devices like kickers or electrostatic septa. The power deposition issue is particularly critical for superconducting devices~\cite{rumolo2001simulation} where the cooling capacity on the beam chamber might be limited (see, e.g. Refs.~\cite{Hatchadourian:364733,braduIPAC16} for some highlights of LHC-related aspects).

Observing the sketch in Fig.~\ref{fig:sketch}, it is possible to identify different factors that influence the EC build-up process. The beam chamber plays an essential role. Its geometry affects the electron acceleration and time-of-flight between impacts, as well as the boundary conditions for the calculation of the electric field generated by the beam and the electrons. The properties of the chamber surface define the amount of electrons generated by photoemission and, more importantly, the probability of secondary emission occurring when an electron impacts on the wall~\cite{PhysRevSTAB.5.124404, PhysRevLett.93.014801}. The secondary emission process is described by the Secondary Electron Yield (SEY) function, which is defined as the ratio between the impinging electron current and the emitted electron current and depends strongly on the energy of the impinging electrons. The SEY depends on the chemical properties of the surface and, for several materials, the SEY decreases when the surface is exposed to an electron flux~\cite{doi:10.1116/1.2049306}. For this reason the EC is, to some extent, a self-curing mechanism in the sense that the surface can be conditioned by exposing it to the EC itself, the so-called beam-induced scrubbing. 

The beam parameters play a role in the EC build-up process and the bunch spacing and the length of the bunch train are key parameters. The bunch intensity and the bunch length also influence the EC dynamics, as they change the force acting on the electrons from the beam, whereas the transverse emittances have a milder impact {\cite{giannithesis}}. This, does not exclude that special transverse beam distributions might have a strong impact on EC phenomena, which is the aspect addressed by our study and profits from the recent studies on the PS Multi-Turn Extraction (MTE)~\cite{MTE-PRL,MTE-PRSTAB1,MTE-PRSTAB2,MTE-prog,MTE-adiab,MTE-EPL,MTE-PRAB,MTE-PRAB1,MTE-fluct}, where transverse beam splitting by means of adiabatic resonance crossing is performed. Indeed, beam splitting can be used to generate multimode beam distributions in the horizontal plane, which {in turn} can be used to manipulate the EC generation {thus contributing to a more efficient surface conditioning}.

The generation of charged particle beams with multimode transverse distribution was pushed with MTE~\cite{MTE-PRL} that was studied at the CERN PS~\cite{MTE-PRSTAB1,MTE-PRSTAB2,MTE-prog,MTE-adiab} and became recently fully operational~\cite{MTE-EPL,MTE-PRAB,MTE-PRAB1,MTE-fluct} for the transfer of proton beams from the PS to the SPS in the framework of the fixed-target physics programme. The original idea~\cite{MTE-PRL} discussed the possibility of splitting a single-Gaussian beam into a multi-Gaussian one by means of particles' trapping into stable islands of phase space. This novel beam manipulation relies on two key points: generation of stable islands in the transverse phase space linked with a non-linear resonance of order $N$; adiabatic crossing of the resonance. The first point can be realised by means of a set of magnets generating non-linear fields, such as sextupoles and octupoles, whereas the second point is achieved by changing slowly the {transverse accelerator tunes}.

Results about yet another domain of application of stable islands are reported here, covering the impact of transverse multimode beam distribution on EC effects (some initial results had been presented in~\cite{MTE_EC_IPAC}). Non-linear beam dynamics allows a single Gaussian distribution to be transformed into multiple Gaussians, which, when projected onto physical space, generate a multimode beam distribution. The properties of such a multimode distribution can be fully controlled so that one can assume to have enough freedom to tailor, almost at will, the transverse beam distribution after splitting. For the case of a single-Gaussian distribution, it is well-known that EC effects are rather insensitive to the transverse beam dimensions {\cite{giannithesis}}: Multimode distributions could be a means to change this feature. Therefore, they could open the possibility to either condition a region of the surface of the beam pipe wider than what could be done with single-Gaussian beams, or to deliver a higher electron dose, which would result in a faster conditioning. In both cases the conditioning process would become more efficient.
\section{Numerical model for generating multimode beams}  \label{sec:model}
To study the proposed beam manipulation and its impact on EC effects, a simplified model of the transverse betatronic motion in a ring, including non-linear effects, has been used. It assumes that the sextupoles and octupoles are located at the starting section of the ring and the ring itself is made of regular cells made of alternating focusing quadrupoles and dipoles in between (also called FODO cells), whose layout is assumed to be that of the SPS~\cite{SPS}. The linear motion is parametrised by the three Twiss parameters~\cite{courant} $\beta_z(s), \alpha_z(s), \gamma_z(s)$, where $z=x, y$ for horizontal or vertical motion, respectively, and $s$ represents the path length from the reference section of the ring. Under the assumption that the non-linear elements are represented as single kicks~\cite{Yellow}, the so-called Poincar\'e map (see Ref.~\cite{Yellow} and references therein), can be written in the form of a polynomial map~\cite{MTE-PRSTAB1}
\begin{equation}
\begin{pmatrix}
{X} \\ {X}'\\ {Y} \\ {Y}'
\end{pmatrix}_{n+1} = \mathcal{R}
 \begin{pmatrix} 
{X} \\ 
{X}'+ \displaystyle{{X}^2 - \chi {Y}^2 +
\kappa \left ({X}^3 - 3 \chi {X} {Y}^2 \right )} \\ 
{Y} \\ 
{Y}'+ \displaystyle{-2 \chi {X} {Y}-\kappa 
\left (\chi^2 {Y}^3 - 3 \chi {X}^2 {Y} \right )} 
\end{pmatrix}_{n}
\label{map4D}
\end{equation}
with $\kappa, \chi$ real parameters, the latter representing the ratio $\beta_y/ \beta_x$ at the location of the sextupoles and octupoles. The map~\eqref{map4D} is a H\'enon-like~\cite{henon} map in 4D, and the components of the vector $({X},{X}',{Y},{Y}')$ are dimensionless coordinates~\cite{Yellow} allowing to set the coefficient of the quadratic term of the map~\eqref{map4D} to one. $\mathcal{R}$ is a 4D matrix of the form $\mathcal{R} =  R(\omega_x) \otimes R(\omega_y)$ with $R(\omega_z)$ a $2\times 2$ rotation matrix
\begin{equation}
R(\omega_z) = 
\begin{pmatrix}
 \phantom{-}\cos{\omega_z} & \sin{\omega_z} \\
          - \sin{\omega_z} & \cos{\omega_z}
\end{pmatrix},
\qquad \omega_z = 2 \pi \nu_z \, . 
\label{rot2d}
\end{equation}
Under the assumption that $\chi \ll 1$ (that can be obtained easily in practice by installing sextupoles and octupole magnets in the lattice at locations where $\beta_y \ll \beta_x$),  the map~\eqref{map4D} can be restricted to the horizontal plane, i.e.,  to $({X},{X}',0,0)$, providing a useful model for our study. 

\begin{figure}[htb]
\centering
\begin{tabular}{@{}cc@{}}
{\includegraphics[trim= 0mm 1mm 0mm 3mm,width=0.47\linewidth,clip=]{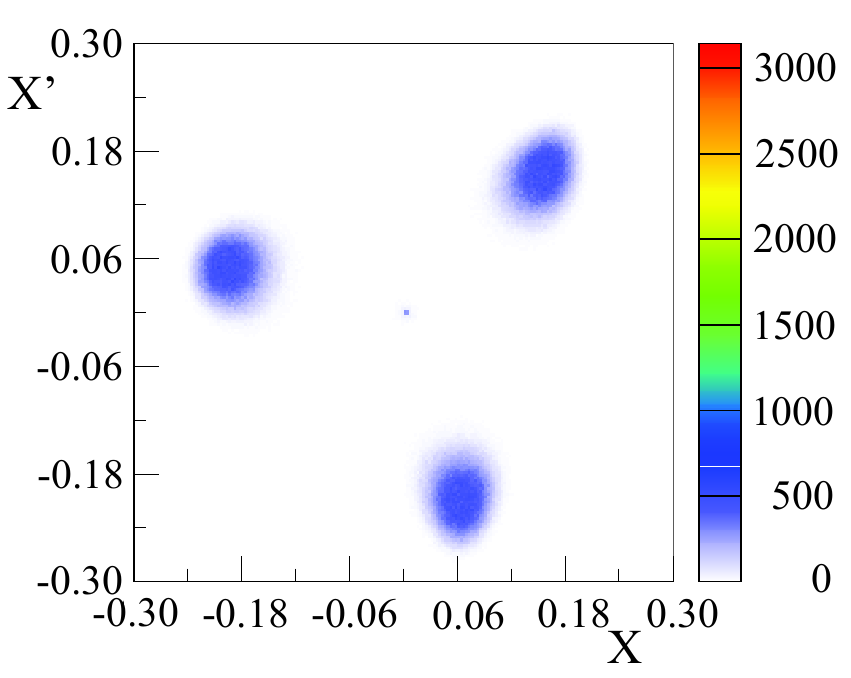}} & 
{\includegraphics[trim= 0mm 1mm 0mm 3mm,width=0.47\linewidth,clip=]{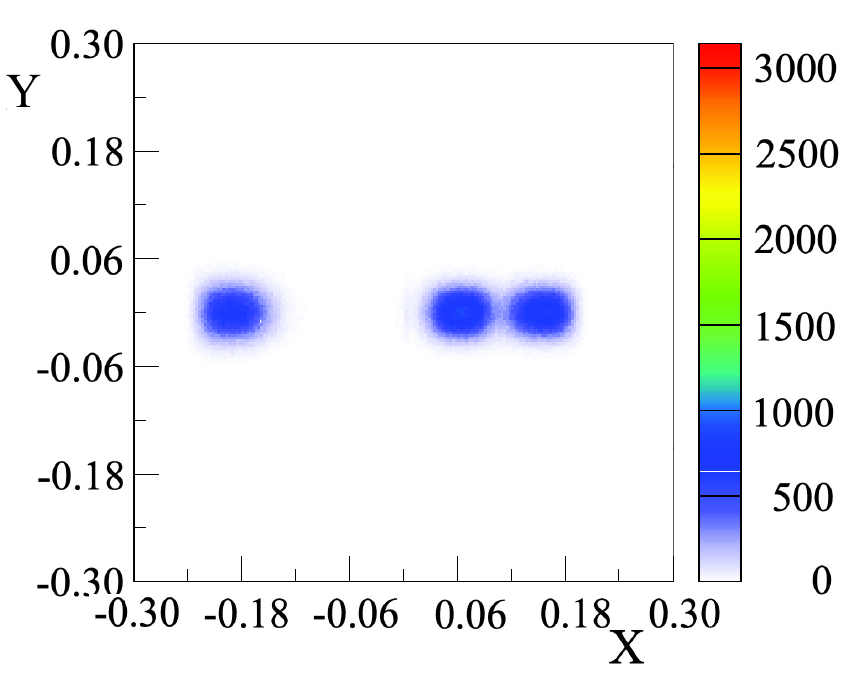}} \\
{\includegraphics[trim= 0mm 2mm 0mm 3mm,width=0.47\linewidth,clip=]{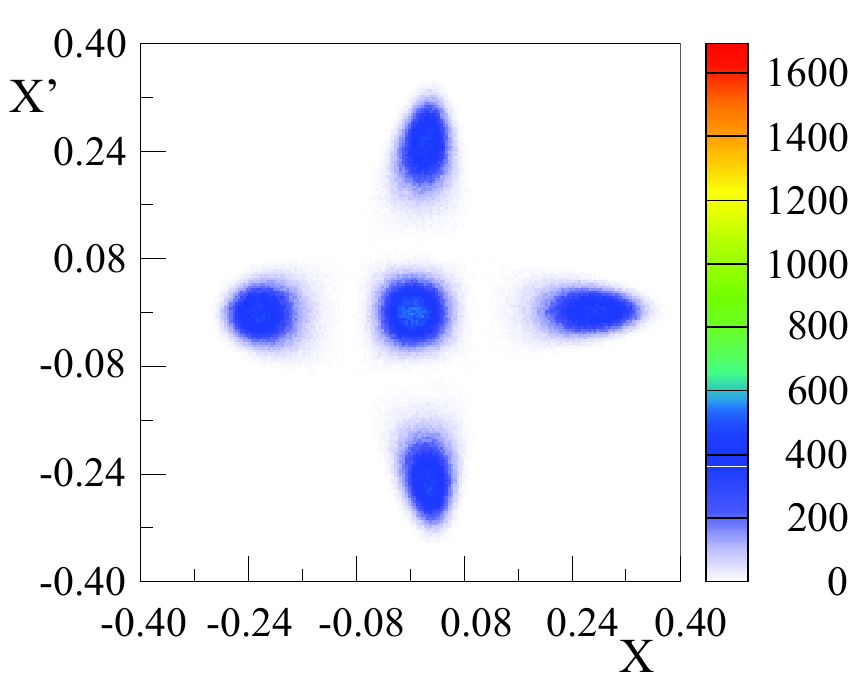}} & 
{\includegraphics[trim= 0mm 2mm 0mm 3mm,width=0.47\linewidth,clip=]{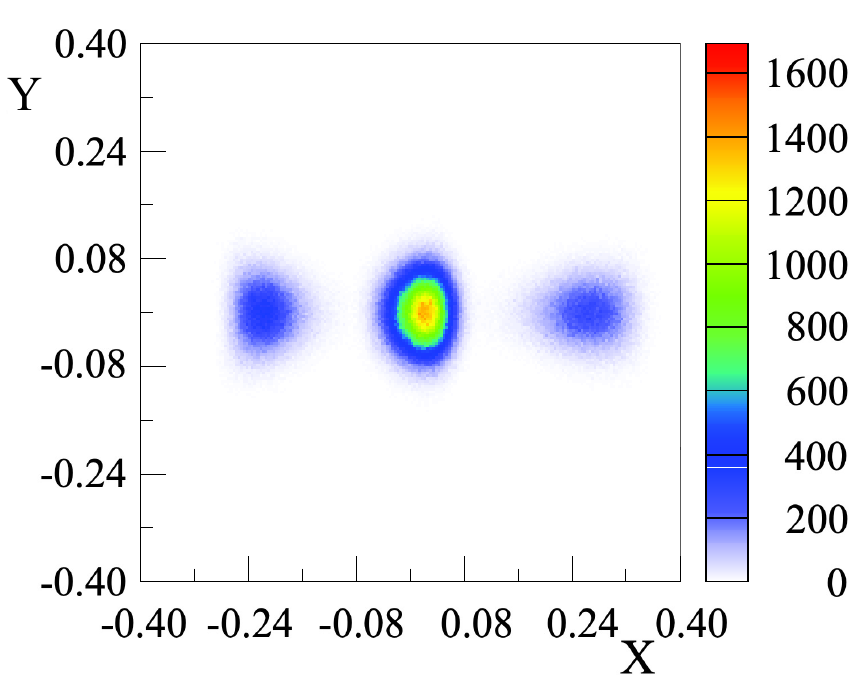}}
\end{tabular}
\caption{Upper: example of split beam obtained with the third-order resonance, phase space (left) and physical space (right). Lower: example of split beam obtained with the fifth-order resonance, phase space (left) and physical space (right).}
\label{splitting}
\end{figure}

The potential benefits of the multi-Gaussian distributions generated by beam splitting should be carefully scrutinised as their evolution as a function of $s$ is intrinsically different with respect to that of single-Gaussian distributions. In fact, assuming the simple circular accelerator model already discussed, the r.m.s. beam size of a single-Gaussian beam varies along the circumference as $\sim \sqrt{\beta_z(s)}$~\cite{courant}. On the other hand, a multi-Gaussian beam follows the position of the fixed points, located inside the stable islands, as they vary along the ring circumference. At the same time, the size of each Gaussian varies as $\sqrt{\hat{\beta}_z(s)}$ where $\hat{\beta}_z(s)$ is the beta-function for the parametrisation of the linear motion around the fixed points, for which in general $\beta_z(s) \neq \hat{\beta}_z(s)$. Finally, for EC effects only the projection of the multi-Gaussian distribution on the horizontal and vertical dimension matters. Therefore, the multimode distribution obtained by projecting the multi-Gaussian one might feature a non-negligible dependence on the position along the ring, in particular in terms of overall width. Hence, the EC effects have been estimated using four transverse distributions: a standard single-Gaussian; a multimode distribution obtained from a three-Gaussian distribution (generated  by crossing the third order resonance); a multimode distribution obtained from a five-Gaussian distribution (generated by crossing the fifth-order resonance); a hollow distribution, expressed in polar coordinates $(r, \theta)$ as 
\begin{equation}
\rho(r, \theta)= \frac{1}{\sqrt{2 \pi} \sigma_r} \, \displaystyle{e^{-\frac{(r-\mu_r)^2}{2 \, \sigma^2_r}}} \, . 
\end{equation}
In Fig.~\ref{splitting} examples of the three-Gaussian  (upper left) and of the five-Gaussian distribution (lower left) are shown in phase space $({X},{X}')$. In the right part, the corresponding multimode distributions are depicted in physical space $({X},{Y})$, with the projection effect clearly visible. The phase-space configurations shown in the left part of Fig.~\ref{splitting} rotate while moving along the ring and at the same time the distance of the islands and their width change, thus changing the projection along ${X}$.

It is clear that the higher the order of the resonance, the higher the number of Gaussians of the split beam and the more regular is the projection as a function of $s$, and the distribution tends to a hollow one, which is the reason why the latter was studied. The single-Gaussian and the hollow distributions represent the low- and high-resonance order limit cases. Parenthetically, the hollow beam distribution is also rather easy to generate in a circular ring, which is an additional advantage and represents yet another argument for including it in our study.

The results have been obtained by generating the four types of distributions as the horizontal beam distributions. For the three- and five-Gaussian cases, the map~\eqref{map4D}, restricted to the horizontal plane, has been used. The vertical distribution has been always assumed to be a single Gaussian. 
\section{Results of numerical simulations of EC effects} \label{sec:results}
\begin{figure}[htb]
\centering
{\includegraphics[trim = 0mm 3mm 0mm 3mm, width=\linewidth,clip=]{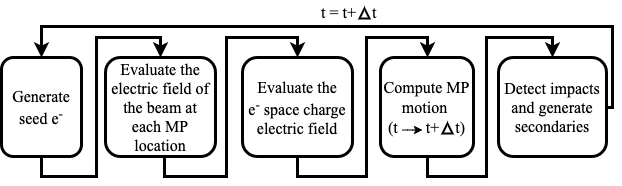}}
\caption{Sketch of the flow diagram of the EC simulations.}
\label{sketch}
\end{figure}

PyECLOUD (see Ref.~\cite{PyEcloud} and references therein) is a 2D macroparticle (MP) code, where the electrons are grouped in MPs to achieve a reasonable computational burden. The dynamics of the MP system is simulated following the flow diagram sketched in Fig.~\ref{sketch}.

At each time step, seed electrons, due to residual gas ionisation and/or to synchrotron radiation-induced photoemission from the chamber walls, are generated with a number consistent with the passing beam slice and with positions and momenta determined by theoretical or empirical models. The electric field acting on each MP is evaluated, i.e. the field of the beam and the space-charge contribution of the electron system. The latter is calculated by a classical Particle in Cell (PIC) algorithm, where the finite difference method is employed to solve the Poisson equation with perfectly conducting boundary conditions on the beam chamber. The knowledge of the total electric field at each MP location allows updating MP positions and momenta by integrating the equations of motion. At this stage, the presence of an external magnetic field can be taken into account. At each time step, the wall hits are detected and a proper model of the secondary emission process is applied to generate charge, energy, and angle of the emitted electrons. Depending on the size of the emitted charge, a rescaling of the impinging MP can be performed or new MPs can be emitted.

To study the effect of multimode beams described in previous sections on the EC formation, numerical simulations were performed for different positions of the various Gaussian distributions for the multi-Gaussian case, and varying $\mu_r, \sigma_r$ for the case of the hollow distribution. This was  complemented by a detailed scan over the horizontal phase advance $\phi_x$. Furthermore, EC simulations have been carried out for both dipolar and quadrupolar external magnetic field. {It is worth mentioning that for the configurations considered in these studies, the longitudinal electrons drift in the quadrupolar field~\cite{Drift} is of the order of a few centimetres, totally negligible with respect to the length of the SPS quadrupoles ($3-3.8$~m). Hence, a 2D treatment is fully appropriate for our purposes.} The key parameters used in the numerical simulations are listed in Table~\ref{parameters}. 
\begin{table}[htb]
\centering
\caption{Parameters used for the EC simulations.}
\label{parameters}
\begin{tabular}{@{}l@{}ccc@{}}
\hline
Parameter                        & Symbol      & Unit    &   Value \\
\hline
Beam momentum                    & $p$         & GeV/{\sl c} & $25.92$ \\
Total bunch intensity            & $N_{\rm b}$ & $10^{11}$ p & $2.5$  \\
Horizontal emittance             & $\epsilon_x$& $\mu$m  & $3$  \\
of single-Gaussian               &             &         &      \\
Vertical emittance               & $\epsilon_y$& $\mu$m  & $3$ \\
of all distributions             &             &         &    \\
External dipolar field           &             & T       & $0.1166$ \\
External quadrupolar field       &             & T/m     & $1.00127$ \\
Elliptical chamber               &             & mm$^2$  & $156\times42.3$ \\
Rectangular chamber              &             & mm$^2$  & $132\times51.5$ \\
\hline
\end{tabular}
\end{table}

{The values of the maximum-SEY parameter (see~\cite{seypaper} for more detail) used in the numerical simulations are in the range $1.4$ to $1.5$, corresponding to the situation observed in the accelerator close to the end of the scrubbing process~\cite{PyEcloud}. This choice is justified by the behaviour of the SEY reduction, which is very fast at the beginning of the scrubbing and much slower at the end of it. Therefore, means to increase the scrubbing efficiency should be looked for in this interval of intermediate SEY values.}

\begin{figure}[!htb]
\centering
{\includegraphics[trim = 1mm 1mm 2mm 2mm, width=0.88\linewidth,clip=]{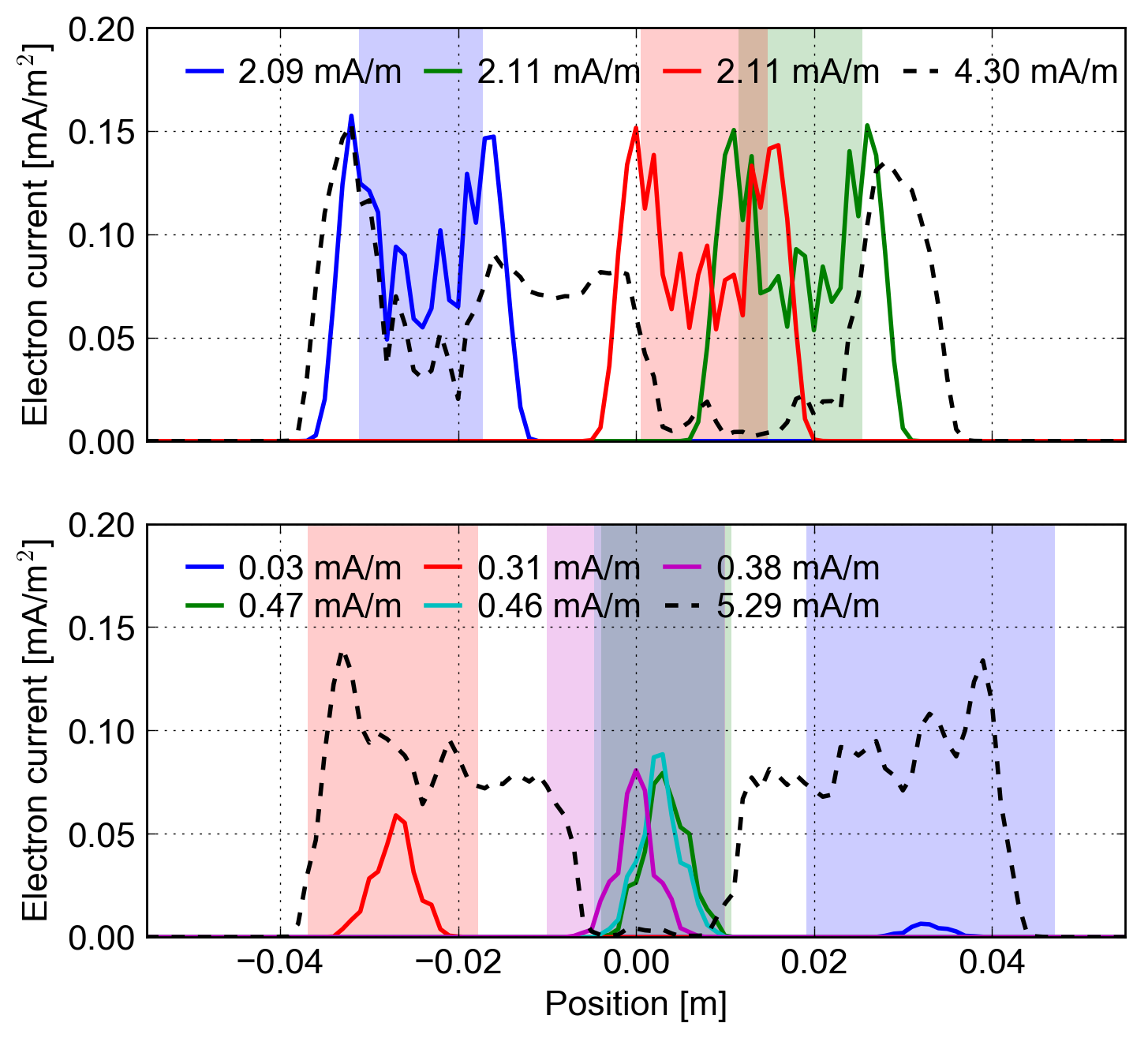}} 
\caption{Distribution of EC current for the three- and five-Gaussian cases (top and bottom, respectively) for an external dipole field, SEY 1.4, rectangular vacuum chamber, and $\phi_x=0$. The coloured lines refer to the EC current distribution for individual Gaussians, the dashed line to the multimode beam case, {the shaded areas to the extent of the projected beam distribution. The reported integrals of the EC current distributions show clearly the difference in EC effects for individual Gaussians or multimode distributions}.}
\label{combination}
\end{figure}

The first aspect investigated is whether the EC generation for multimode distributions can be derived from that of each individual Gaussian distribution. Numerical simulations have been performed for both three- and five-Gaussian distributions as well considering each single Gaussian individually and the results are shown in Fig.~\ref{combination}.

In the top row, the three-Gaussian distribution case is reported and the interesting observation is that indeed there is a non-linear interaction between the three Gaussians and the EC current distribution, so that the total number of electrons for the multimode case is different with respect to a linear addition of the results for individual Gaussians. This is even more striking by considering the case for the five-Gaussian beam (bottom row). The same non-linear behaviour is observed, but here there is also a non-negligible  increase of the number of produced electrons for the multimode distribution with respect to the single-Gaussian case. {In both cases, the EC distribution generated by the multimode distribution is wider and more symmetrical than that obtained by superposing the individual EC distributions. This is due to the fact that the EC generation is driven by the superposition of all electric fields and hence it is less sensitive to small fluctuations in the properties of the individual Gaussians when these are present simultaneously.} 

\begin{figure*}[htb]
\centering
\includegraphics*[trim = 0mm 18mm 0mm 7mm, width=0.45\textwidth,clip]{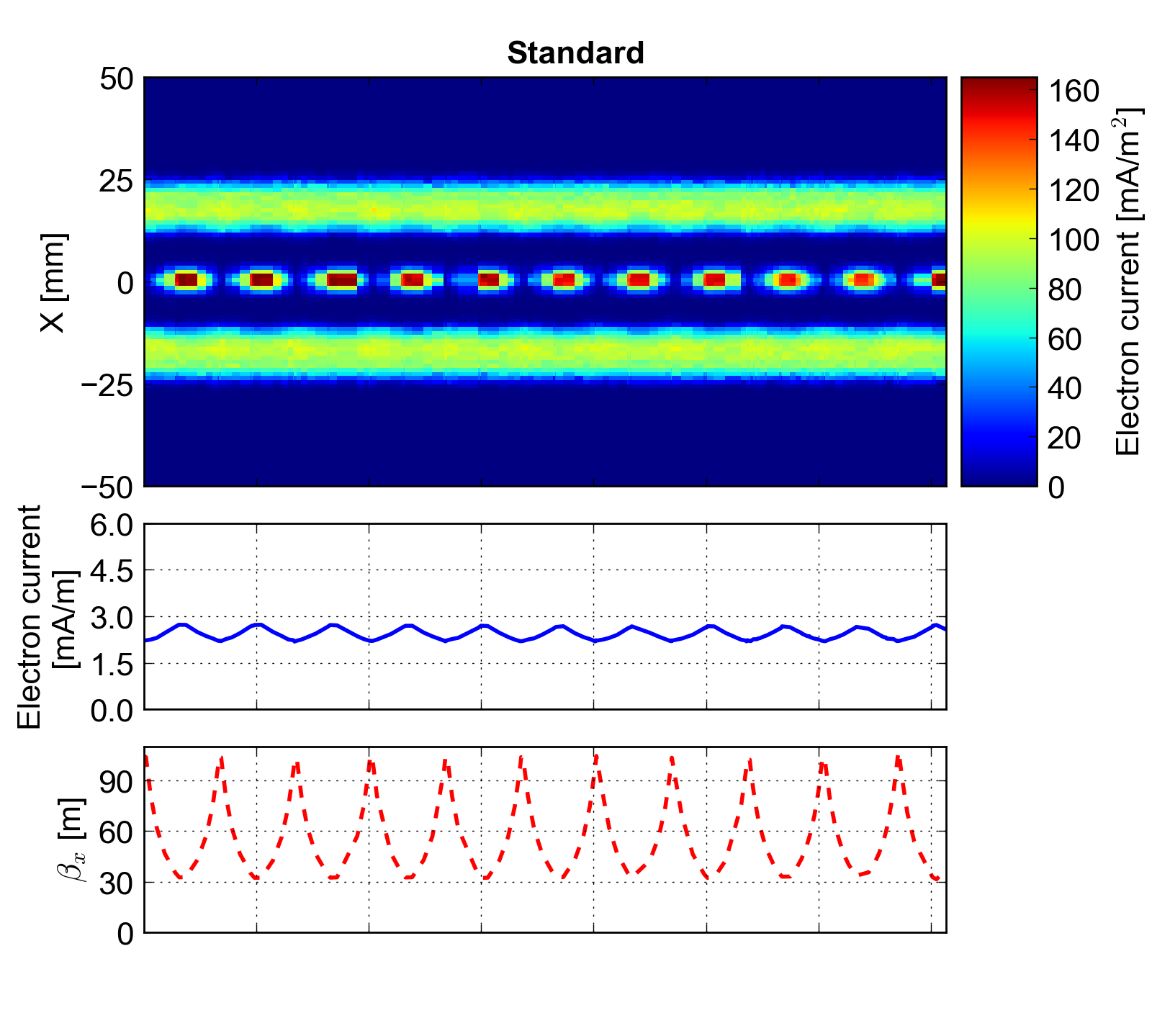}
\includegraphics*[trim = 0mm 18mm 0mm 7mm, width=0.45\textwidth,clip]{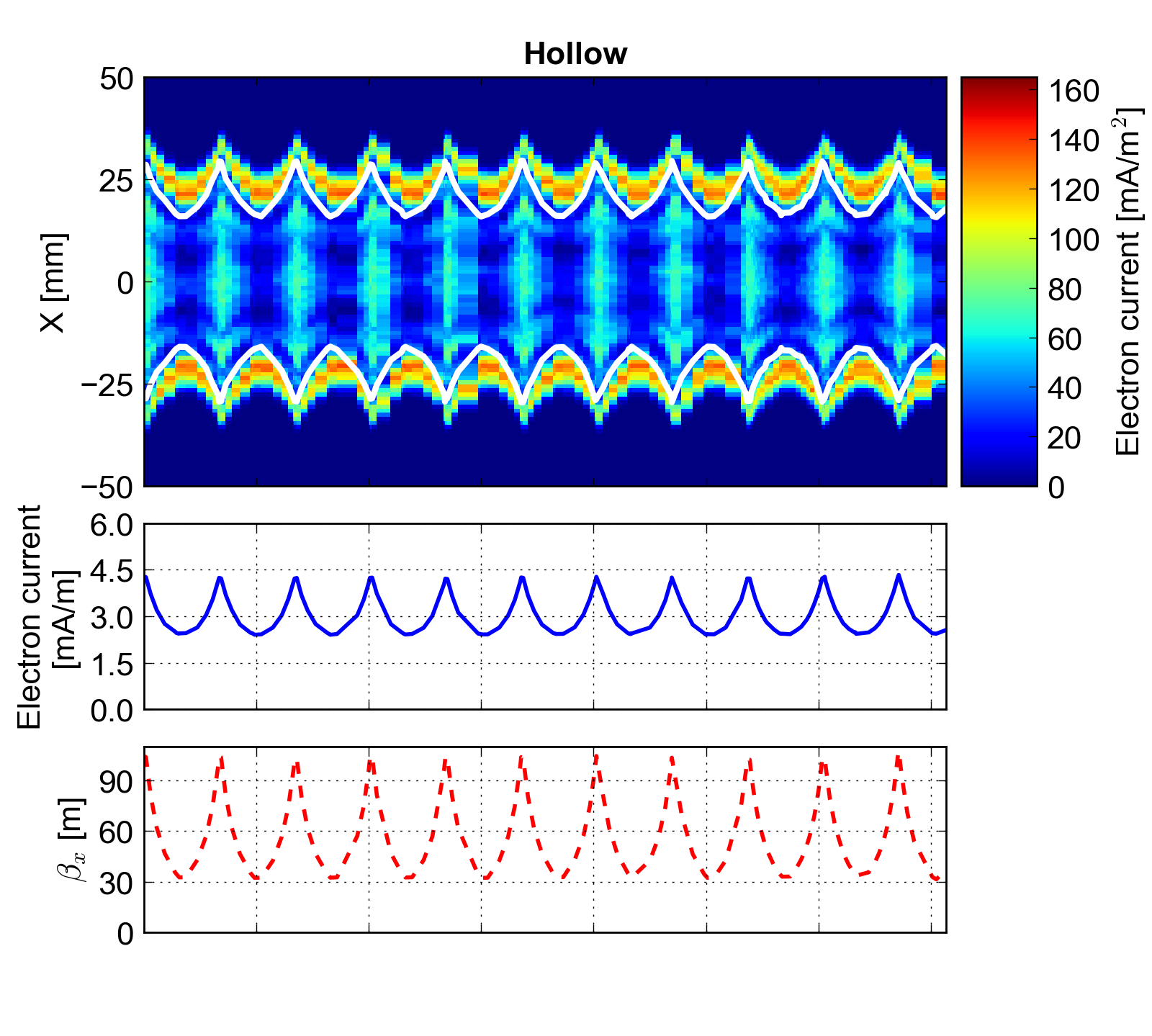}\\
\vspace{2mm}
\includegraphics*[trim = 0mm 2mm 0mm 2mm, width=0.45\textwidth,clip]{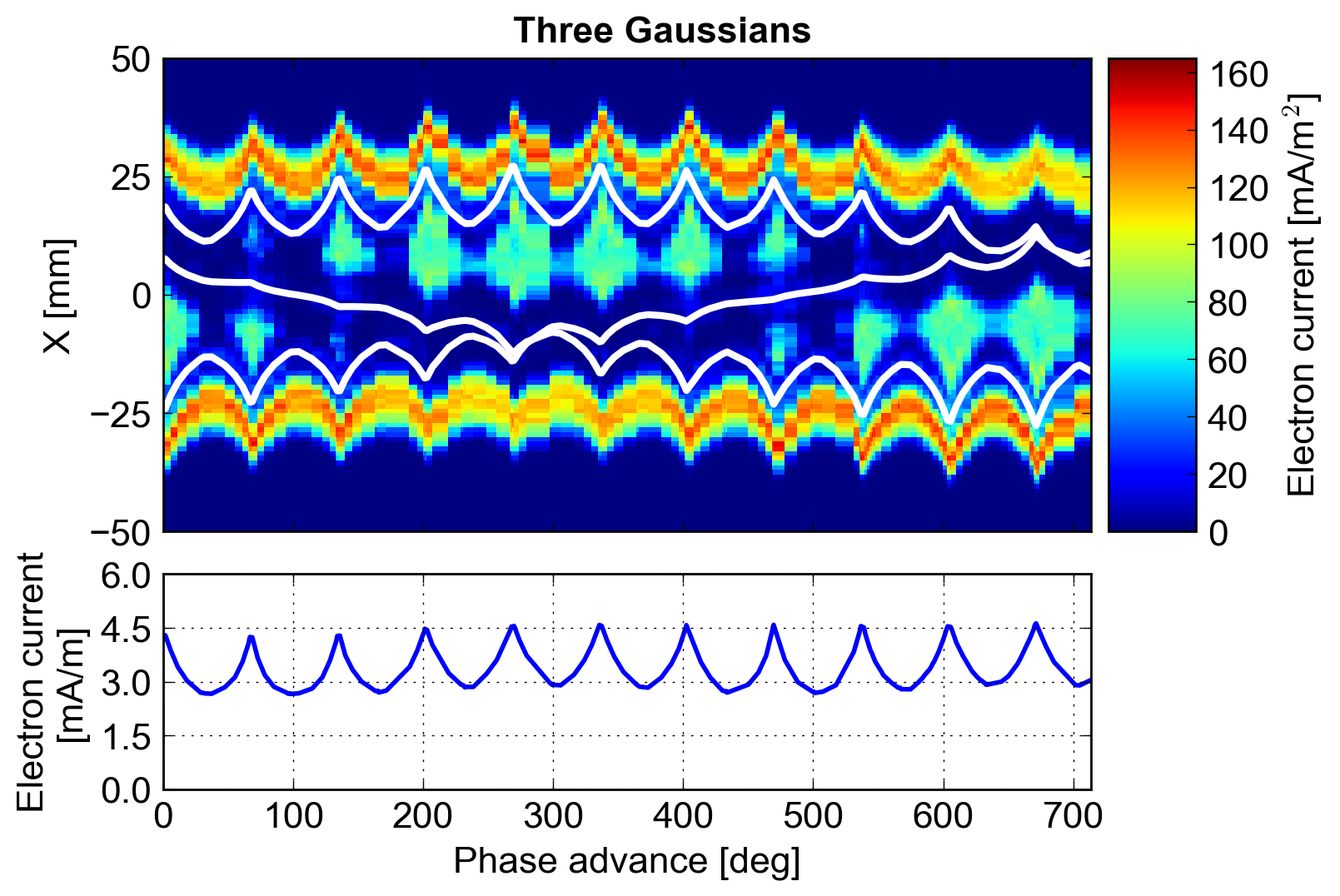}
\includegraphics*[trim = 0mm 2mm 0mm 2mm, width=0.45\textwidth,clip]{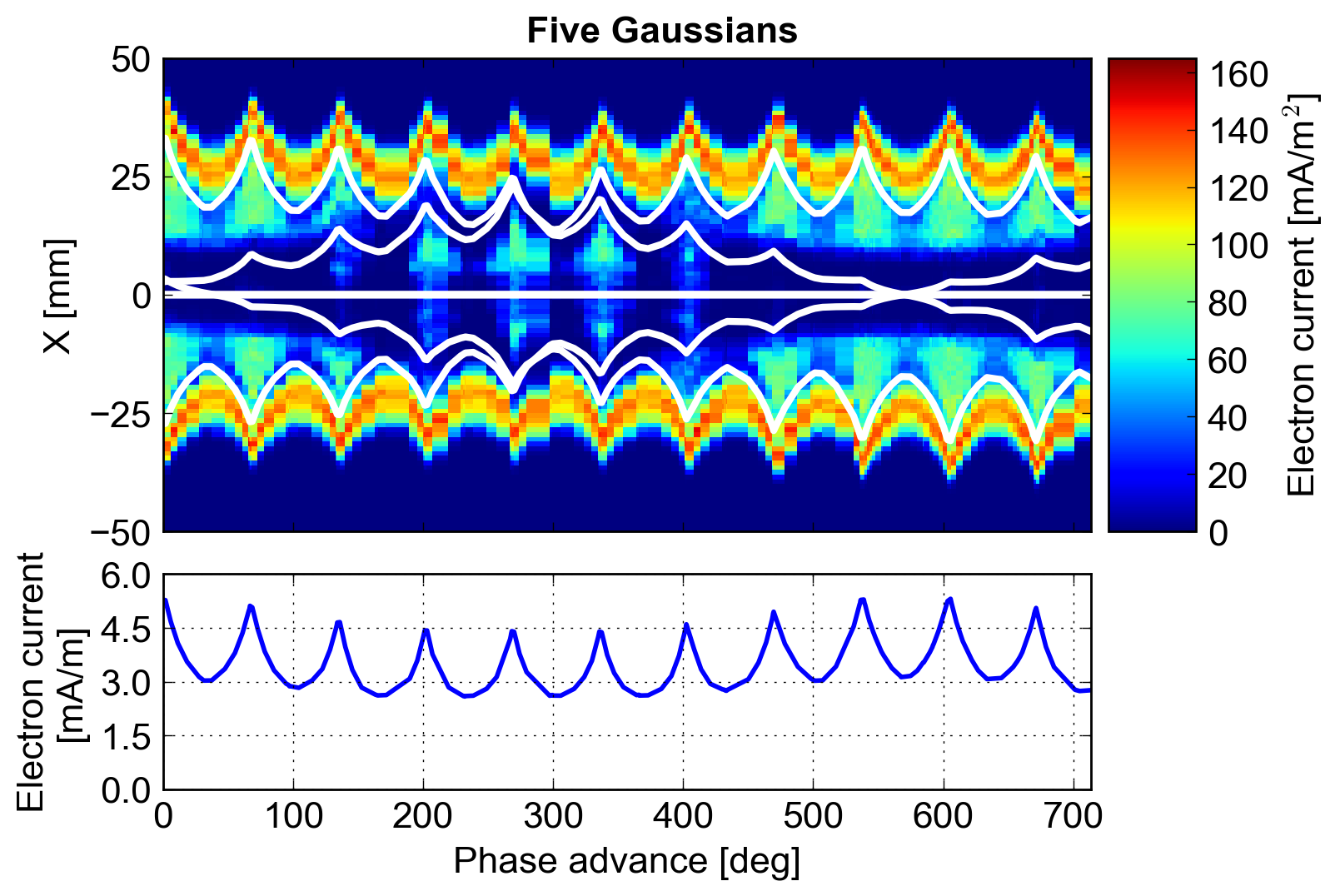}
\caption{Horizontal distribution of the EC current for SEY 1.4 (colour plot, the white lines represent the peaks of the hollow, three- and five-Gaussian beams), integrated EC current (blue line), and $\beta$-function (dashed red line) as a function of $\phi_x$ for for the case of a dipole field and rectangular vacuum chamber. Standard, hollow, and multimode (three- and five-Gaussian) beams are shown. The EC current is strongly enhanced with respect to the standard beam and modulated in correlation with $\beta_x$.}
\label{dipole}
\end{figure*}

The results of extensive numerical simulations to determine the dependence on the horizontal phase advance $\phi_x$, for a dipolar external field, are shown in Fig.~\ref{dipole}. The plots represent the behaviour of horizontal distribution of EC current, EC current, and $\beta_x$ as a function of $\phi_x$. The four types of beam (single-Gaussian, hollow, three- and five-Gaussian) are shown. The white lines indicate the position of the peaks of the projected beam distributions for the hollow, three- and five-Gaussian cases. 

Higher densities for EC current can be obtained with multimode distributions, which is clearly visible from the density plots. A stronger modulation of the EC current density is also observed for the multimode distributions, correlated with the modulation of the $\beta$-function. While the characteristic constant stripes of the EC distribution are visible for the standard, single-Gaussian beam, these are replaced by more complex structures, whose width and position are changing as a function of $\phi_x$. The plots of the EC currents clearly reveal the presence of the $\beta$-wave. The minimum EC current for the multimode is comparable with the single-Gaussian case, while the maximum EC current exceeds that of the single-Gaussian beam. 

\begin{figure*}[htb]
\centering
\includegraphics*[trim = 0mm 18mm 0mm 7mm, width=0.45\textwidth,clip]{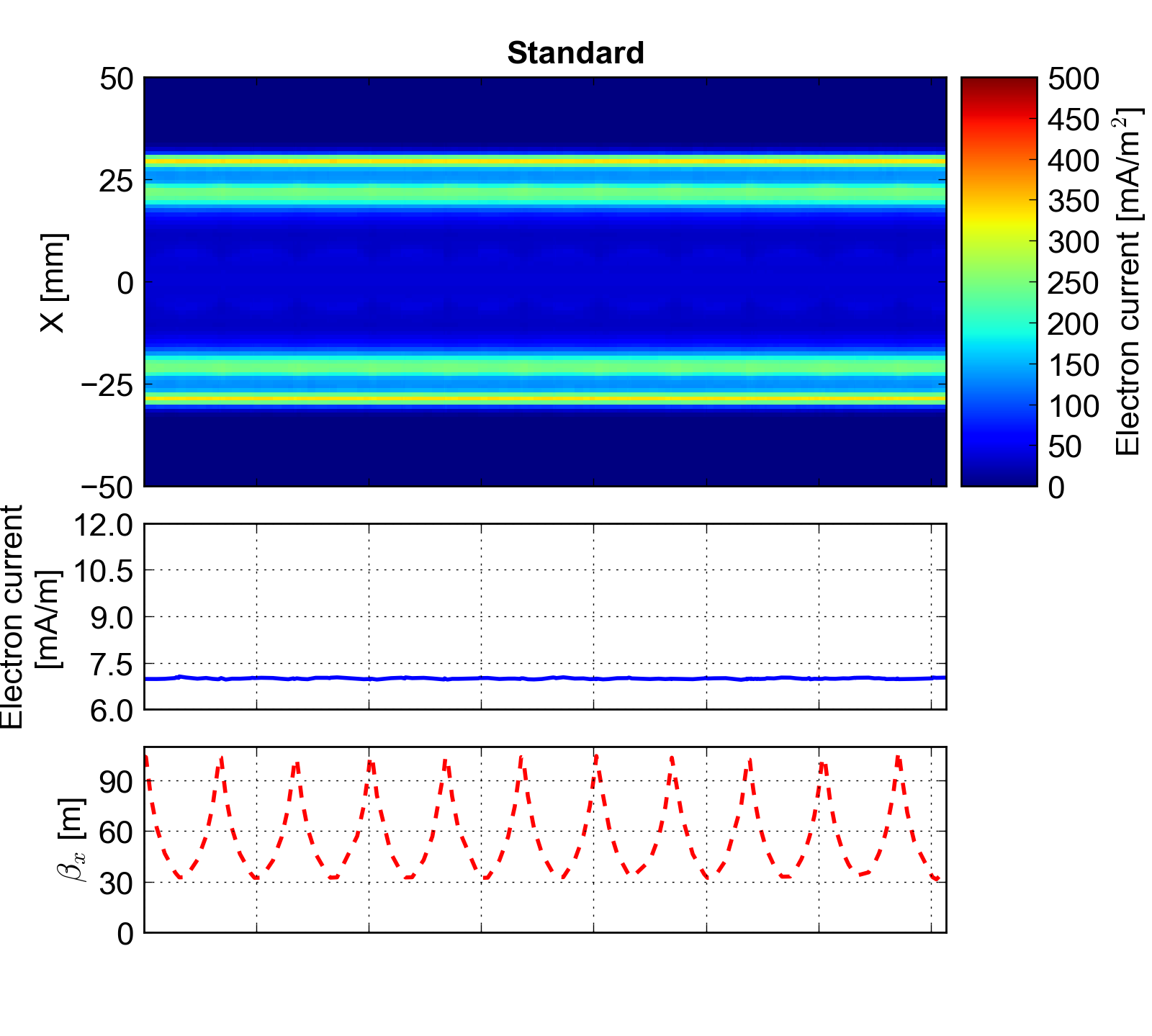}
\includegraphics*[trim = 0mm 18mm 0mm 7mm, width=0.45\textwidth,clip]{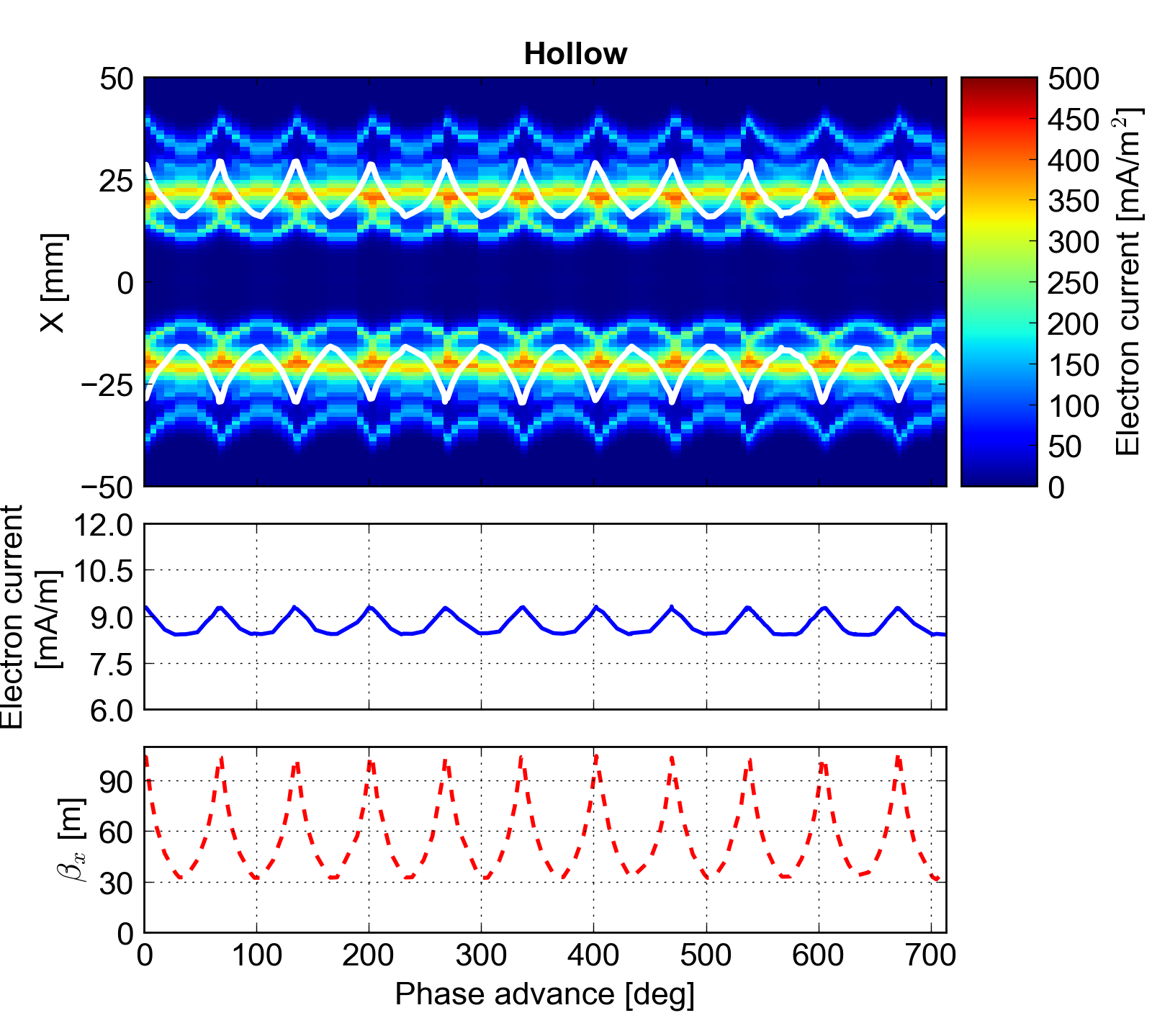}\\
\vspace{2mm}
\includegraphics*[trim = 0mm 2mm 0mm 2mm, width=0.45\textwidth,clip]{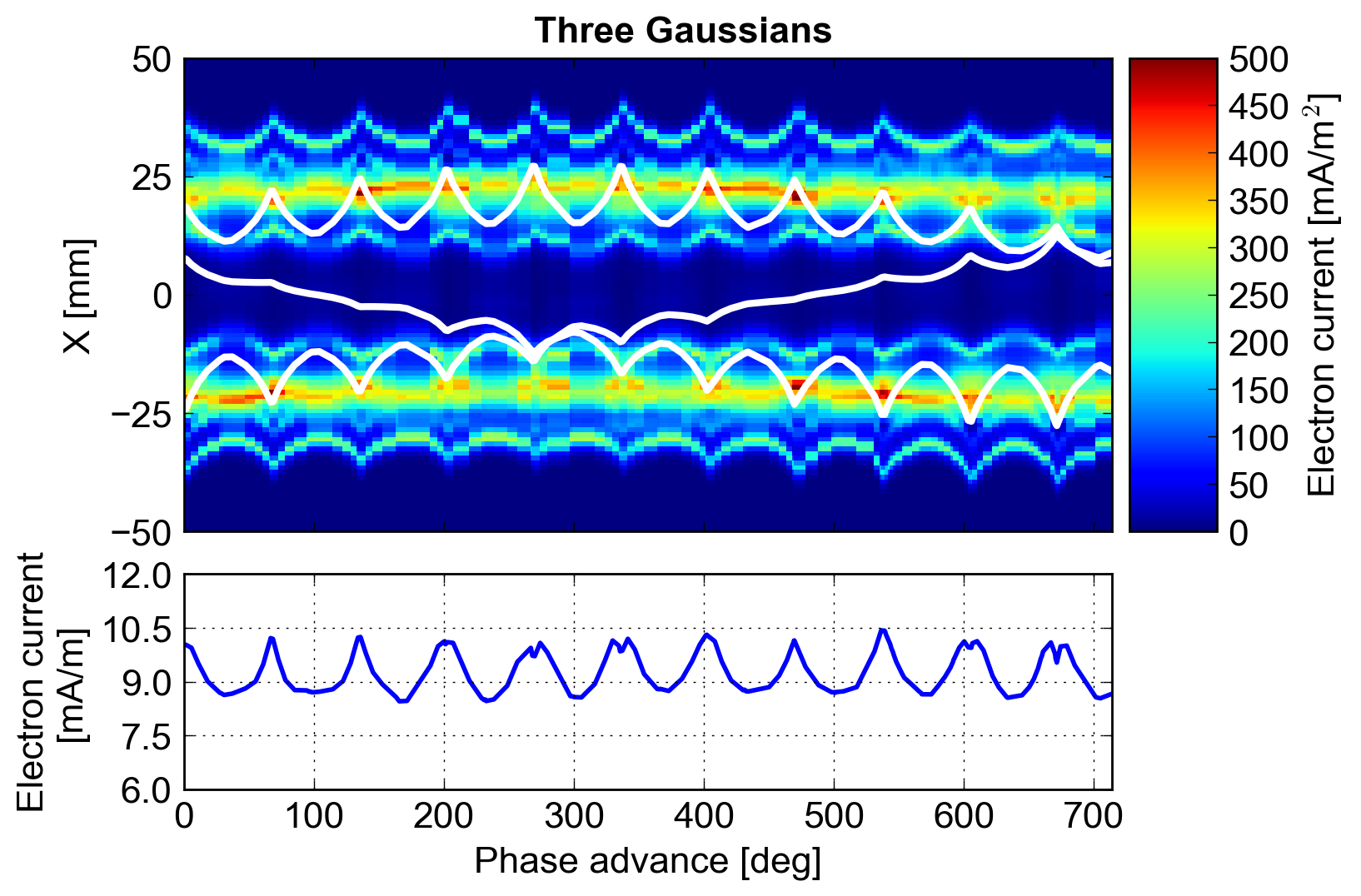}
\includegraphics*[trim = 0mm 2mm 0mm 2mm, width=0.45\textwidth,clip]{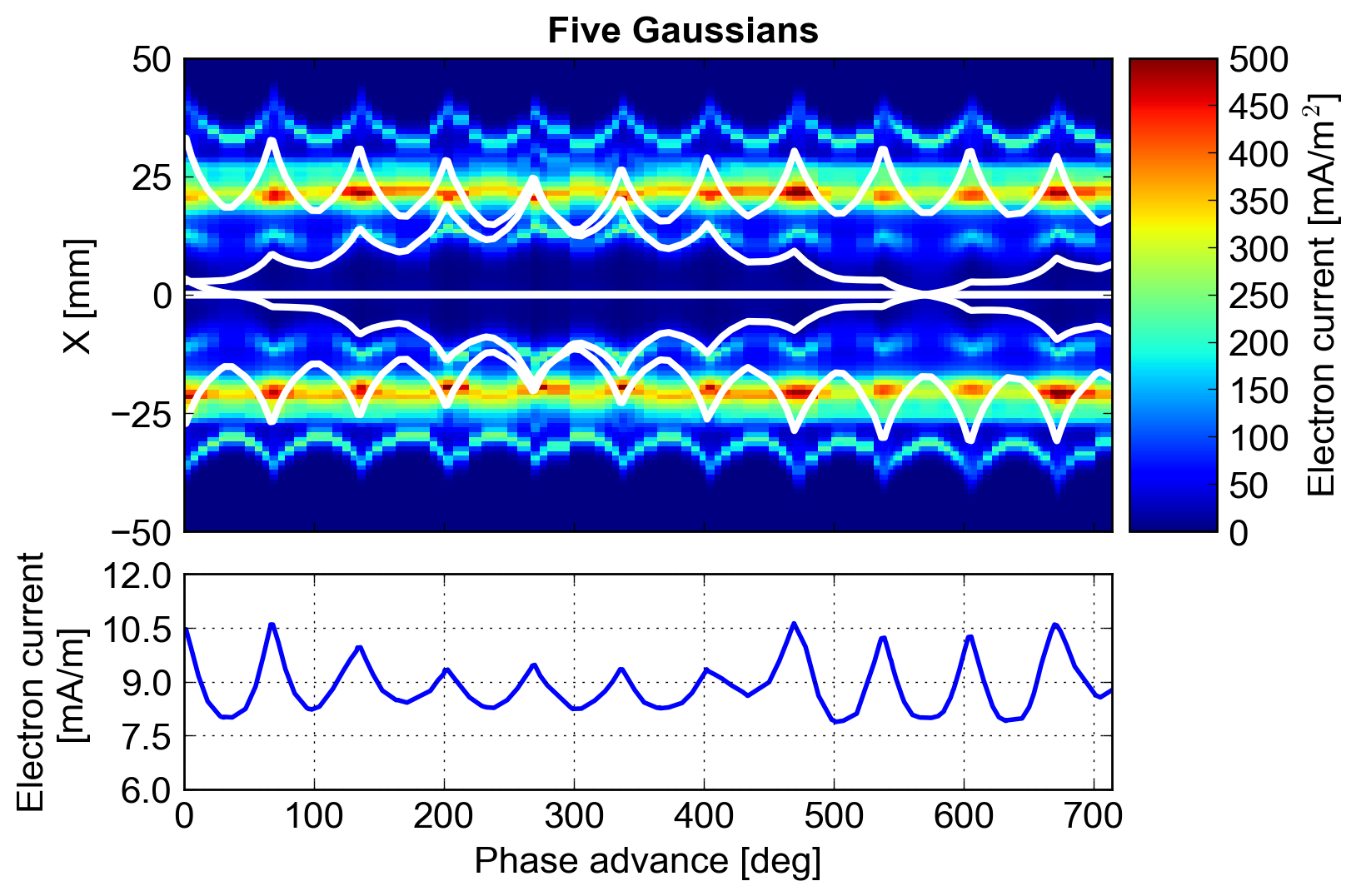}
\caption{Horizontal distribution of the EC current for SEY 1.5 (colour plot, the white lines represent the peaks of the hollow, three- and five-Gaussian beams), integrated EC current (blue line), and $\beta$-function (dashed red line) as a function of $\phi_x$ for the case of a quadrupole field and elliptical vacuum chamber. Standard, hollow, and multimode (three- and five-Gaussian) beams are shown. The EC current is strongly enhanced with respect to the standard beam and modulated in correlation with $\beta_x$.}
\label{quadrupole}
\end{figure*}

The results of the numerical simulations performed in the presence of an external quadrupolar field are shown in Fig.~\ref{quadrupole}. The considerable increase of the EC current density with respect to the case with external dipole field is visible. Also in this case, the EC current features a non-negligible modulation as a function of $\phi_x$ and, consequently, of $\beta_x$ for the multimode beam distributions. Unlike the results shown in Fig.~\ref{dipole}, the minimum EC current {for all the} multimode beam distributions is always considerably larger than the current for the single-Gaussian case, which indicates a strong enhancement of the EC effects by means of the multimode distributions. 
Among the multimode distributions, the hollow one features a more regular behaviour as function of $\phi_x$ compared to the three- or five-Gaussian ones, no matter the type of external magnetic field. This is due to the more regular distribution of the projection on the physical space. 

\begin{figure}[htb]
\centering
\includegraphics[trim = 8mm 2mm 3mm 2mm, width=0.7\linewidth,clip]{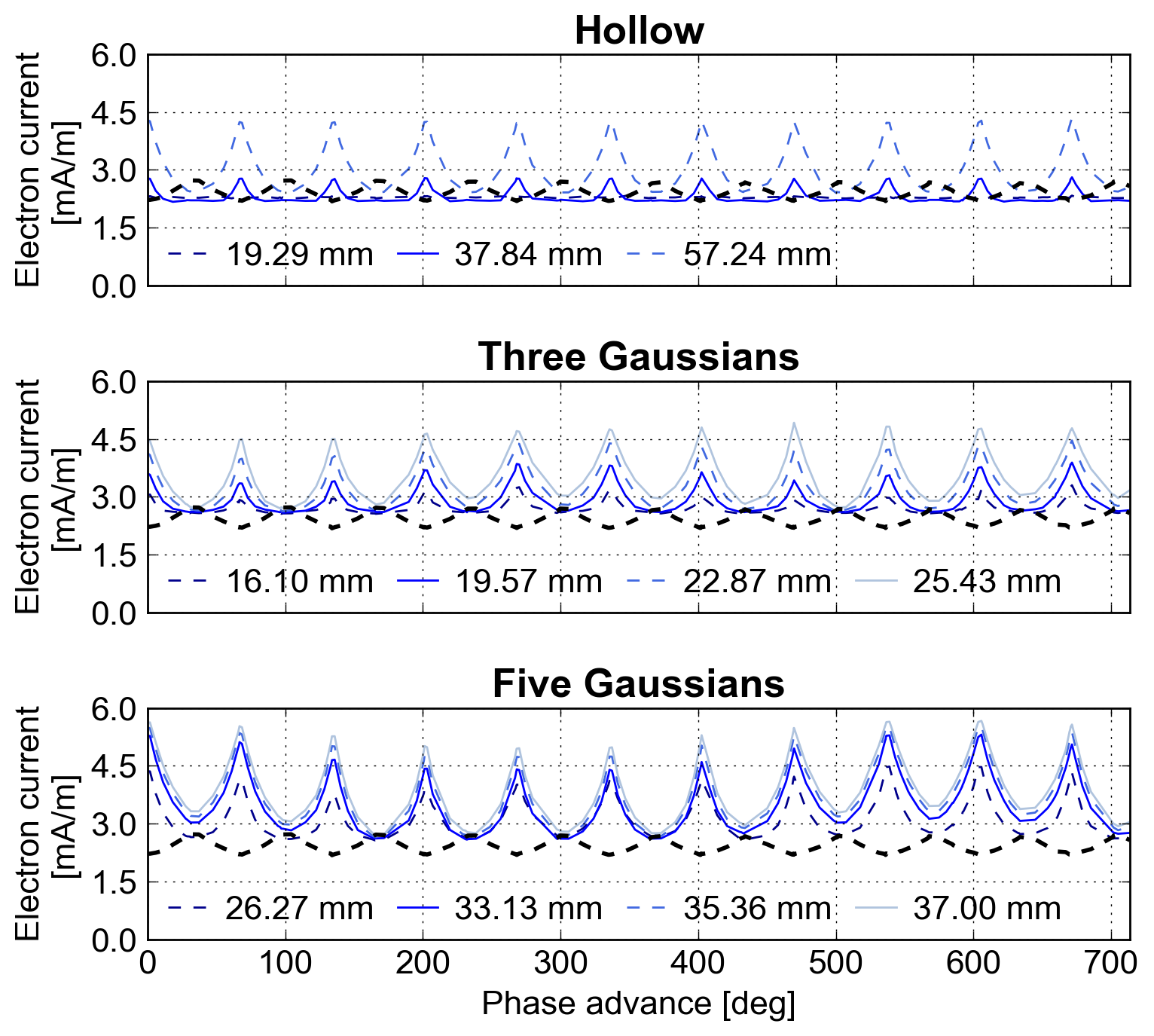} \\
\includegraphics[trim = 8mm 2mm 3mm 2mm, width=0.7\linewidth,clip]{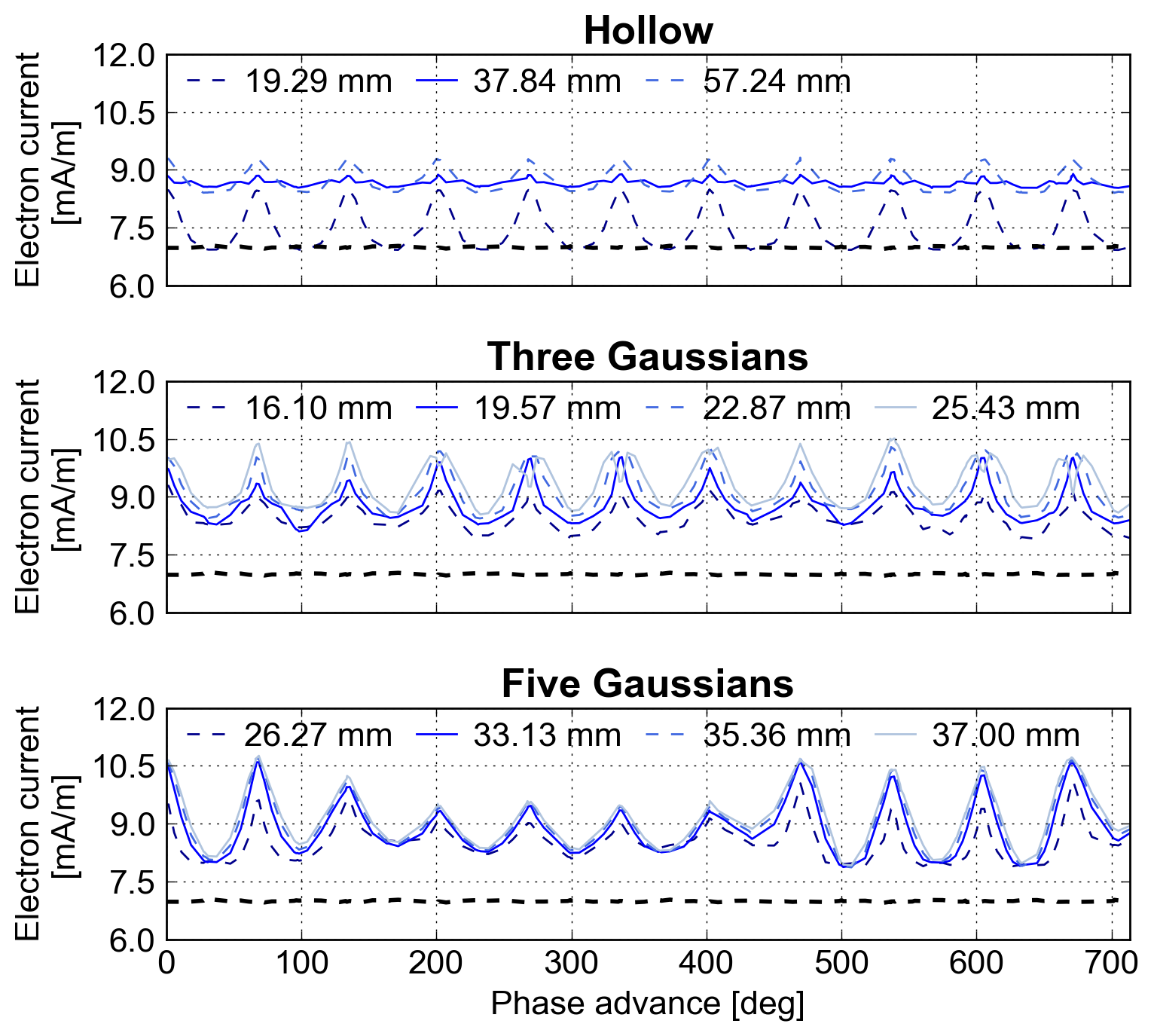}
\caption{EC current as a function of $\phi_x$ for different {values of the maximum absolute transverse} positions of the peaks of the multimode distributions. The black dashed line represents the EC current generated by the single-Gaussian beam. In the first and second block of three plots the dipole and quadrupole external field cases are shown, respectively. {The other parameters are equal to those used for the data shown in Figs.~\ref{dipole} and~\ref{quadrupole}.}}
\label{amplitude}
\end{figure}

From the results shown in Figs.~\ref{dipole},~\ref{quadrupole} it is clear that the EC generation depends on the position of the peaks of the multimode distributions, meaning that both their amplitude and phase at a given location around the ring matter. The {transverse position of the Gaussians} has been varied {in phase space by acting on $\omega_x$} to probe the dependence of the EC current {on such a parameter, while all others are kept constant} and the results are shown in Fig.~\ref{amplitude}.

The curves representing the EC current for several amplitudes of the peaks of the multimode distributions are plotted, together with the current generated by a single-Gaussian beam used as a reference case, for the cases corresponding to dipole and quadrupole external fields. In general, the three- and five-Gaussian cases are always more favourable in terms of EC current than the reference case. The EC current is particularly enhanced when the external magnetic field is of quadrupolar type. As far as the hollow distribution is concerned, a strong enhancement of the EC current is observed for specific values of the amplitude of the peaks. Therefore, while the three- and five-Gaussian distributions {seem to outperform} the single-Gaussian one over a wide range of parameters, the Hollow distribution can become a very appealing {alternative to} the single-Gaussian for specific values of the parameters (mainly the amplitude of the peaks), also considering that it can be very easily generated by means of beam filamentation together with the effect of external non-linearities. 

It is worth noting that the dependence of the EC current is not monotonous with respect to the amplitude of the peaks of the multimode distributions and an optimum amplitude exists that maximises the EC current. This indicates that {the optimal choice} of the position of the peaks of the multimode distributions is a complex problem, involving not only the beam properties, but also the vacuum chamber geometry (and material). However, the extended parameter space ensures that the solution is, in general, {more efficient than} with a simple single-Gaussian beam.
\section{Conclusions} \label{sec:conclusions}
Novel and promising results about EC effects in the presence of multimode-distribution beams have been presented in this paper. EC current can be enhanced by these special beams with respect to single-Gaussian ones. A strong non-linear interaction between the individual Gaussians makes these multimode distributions very efficient in EC generation. The dependence of EC effects on several features of the multimode distributions has been studied and the impact of the change in the projected distributions along the ring circumference has been assessed. Multimode distributions remain superior to single-Gaussian ones in spite of the variation of EC effects along the accelerator circumference. The extremely encouraging results are obtained both in presence of a dipolar or quadrupolar external field, thus making the proposed multimode beams a concrete option to mitigate EC effects by surface conditioning. It is worthwhile stressing that {stronger EC effects imply stronger impact on beam quality, due, e.g. to beam instabilities. Therefore, during the dedicated runs for surface conditioning, the parameters controlling the properties of the multimode distributions have to be varied in order to maintain the beam always slightly below the instability threshold, thus ensuring an optimal performance of the process. It is also worth mentioning that} the hollow beam distribution turns out to be also very effective in enhancing EC phenomena. Future investigations will be devoted to the detailed understanding of the interesting and useful features unveiled during this study in order to further improve the understanding, control, and exploitation of EC effects in presence of multimode beam distributions. 
\end{document}